\documentclass[%
]{revtex4-1}

\usepackage{graphicx}
\usepackage{dcolumn}
\usepackage{bm}
\usepackage{float}
\usepackage{rotating}

\begin{document}
\maxdeadcycles=100000

\begin{center} 
{\bf\Large Satellites of the Broucke-Hadjidemetriou-H\'{e}non family of periodic unequal-mass three-body orbits}

\vspace{0.5cm}

Xiaoming Li$^{1,2}$ and  Shijun Liao$^{3,4*}$ \\ 
\vspace{0.25cm}
$^1$MOE Key Laboratory of Disaster Forecast and Control in Engineering,  School of
Mechanics and Construction Engineering, Jinan University, Guangzhou 510632,
China\\
$^2$Department of Earth, Atmospheric and Planetary Sciences,
Massachusetts Institute of Technology, \\
Cambridge, Massachusetts 02139, USA\\
$^3$ School of Physics and Astronomy,  Shanghai Jiaotong University, Shanghai 200240, China\\  
$^4$ Center of Advanced Computing, School of Naval Architecture, Ocean and Civil Engineering, Shanghai Jiaotong University, Shanghai 200240, China \\ 
$^*$Corresponding author:  sjliao@sjtu.edu.cn

\vspace{0.cm}

\end{center}
 {\bf Abstract}
{\em

The Broucke-Hadjidemetriou-H\'{e}non's  (BHH) orbits are a family of periodic orbits of the three-body system with the simplest topological free group word $a$, while the BHH satellites have free group words $a^k$ ($k>1$), where $k$ is the topological exponent.  Jankovi\'{c} and Dmitra\v{s}inovi\'{c} [Phy. Rev. Lett. 116, 064301 (2016)]  reported 57 new BHH satellites with equal mass and found that at a fixed energy  the relationship between the angular momentum ($L$) and the topologically rescaled period ($T/k$) is the same for both of the BHH orbits ($k=1$) and  the BHH satellites ($k>1$).    
In this letter, we report 419,743 new BHH orbits ($k=1$) and 179,253 new BHH satellites ($k>1$) of the three-body system  with unequal mass, which have never been reported, to the best of our knowledge.  Among these newly-found 598,996 BBH  orbits and satellites,  about 33.5\% (i.e.,  200,686) are linearly stable and thus many among them might be observed in practice.   Besides,  we discover that, for the three-body system with unequal mass at a fixed energy,  relationship between the angular momentum ($L$) and topologically rescaled period ($T/k$) of the BHH satellites ($k>1$) is different from that  of the BHH orbits ($k=1$).  
}

\vspace{0.5cm}

\section{Introduction}
The three-body problem can be traced back to Newton in 1680s, but is still an open question in astrophysics and mathematics today, mainly  because it is not an integrable system  \cite{Musielak2014} and besides has the sensitivity dependance on initial condition (SDIC) \cite{Poincare1890} that broke a new field of scientific research, i.e. chaos.  Even today the three-body problem is still one of central issues for scientists \cite{Stone2019}.    Especially,  periodic orbits of triple system play an important role since they  are 
``the only opening through which we can try to penetrate in a place which, up to now, was supposed to be inaccessible'', as pointed out by Poincar\'{e} \cite{Poincare1890}.   However,  since the famous three-body problem was first put forward, only three families of periodic orbits were found in about three hundred years: (1) the Lagrange-Euler family discovered by Lagrange and Euler in the 18th century; (2) the Broucke-Hadjidemetriou-H\'{e}non (BHH) family \cite{Broucke1975a,  Hadjidemetriou1975a,  Henon1976}; (3) the figure-eight family, discovered numerically by Moore \cite{More1993} in 1993 and then proofed by Chenciner \& Montgomery \cite{Chenciner2000} in 2000,     
 until 2013 when  \v{S}uvakov and Dmitra\v{s}inovi\'{c} \cite{Suvakov2013} numerically found 13 distinct periodic orbits of the three-body system with equal mass.   In recent years, numerically searching for periodic orbits of the three-body system has been received much attention  \cite{Suvakov2014a,  Li2017-SciChina,  Li2018,  Dmitravsinovic2018}.    
\v{S}uvakov \cite{Suvakov2014a} reported the satellites of the figure-eight periodic orbit with equal mass.   Especially,  more than six hundred new families of periodic orbits of equal-mass three-body system were found by Li and Liao \cite{Li2017-SciChina} using a new numerical strategy, namely the clean numerical simulation (CNS) \cite{Liao2009, Liao2014-SciChina, Hu2020} that can give the convergent/reliable numerical solution of chaotic systems in a long enough duration.    Li et al. \cite{Li2018} further used the CNS to obtain more than one thousand new families of periodic orbits of three-body system with two equal-mass bodies.  All of these greatly enrich our knowledge of the famous three-body problem.

With the topological classification method \cite{Montgomery1998},
the BHH orbits have the simplest topology (free group word $w=a$), while the BHH satellites have more free group words $w=a^k$, where $k$ is the topological exponent.  Recently, Jankovi\'{c} and Dmitra\v{s}inovi\'{c}  \cite{Jankovic2016}  reported 57  BHH satellites ($k>1$)  with {\em equal} mass.   Especially,  it was found \cite{Jankovic2016}  that the relationship between the scale-invariant angular momentum ($L$) and the topologically rescaled period ($T/k$) is  the {\em same} for {\em both} of the BHH orbits ($k=1$) and satellites ($k>1$).    However,  all of these  BHH  orbits and satellites have {\em equal} mass. The BHH satellites with {\em unequal} mass have never been reported yet, to the best of our knowledge.  

In this letter, we  investigate the BHH orbits ($k=1$) and satellites ($k>1$)  with {\em unequal} mass.  In \S~2,  the basic ideas of the numerical continuation method are briefly introduced.   In \S~3, we numerically search for the BHH orbits and satellites with unequal mass, and show the corresponding results.   Brief conclusions are given in \S~4.

\section{The initial configuration and numerical method}
\begin{figure*}[]
  \centering \includegraphics[scale=0.25]{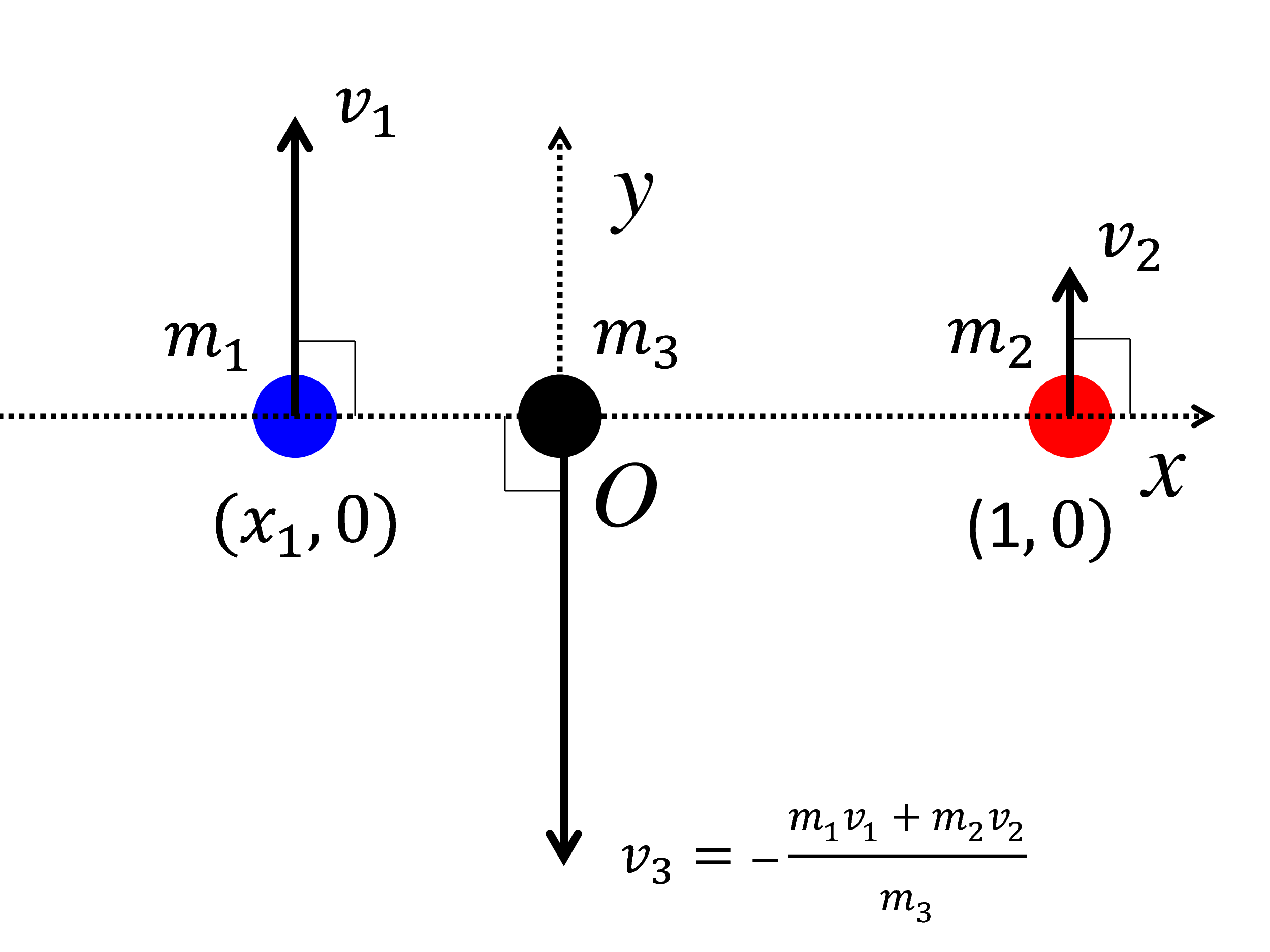}
  \caption{(color online.) The initial configuration of the three-body system.  Here $m_1, m_2$ and $m_3$ denote the mass of body-1, body-2 and body-3, respectively. The corresponding initial velocities are $v_1, v_2$ and $v_3=-(m_1 v_1+m_2 v_2)/m_3$, and the corresponding initial positions are $(x_1,0)$, (1,0) and  (0,0).}
  \label{fig1}
\end{figure*}

Let us consider a three-body system in the Newtonian gravitational field.
Without loss of generality,  let the Newtonian gravitational constant $G=1$.
As shown in Figure~\ref{fig1}, 
the three bodies have collinear initial configuration for the BHH family of periodic orbits:
$\bm{r}_1(0)=(x_1,0)$,  $\bm{r}_2(0)=(x_2,0)$,   $\bm{r}_3(0)=(x_3,0)$, 
and their initial velocities are orthogonal to the  line determined by the three bodies:  
 $\dot{\bm{r}}_1(0)=(0,v_1)$,  $\dot{\bm{r}}_2(0)=(0,v_2)$, $\dot{\bm{r}}_3(0)=(0, v_3)$.

Due to the homogeneity of the potential field of the three-body system, there is a scaling law  : $\bm{r'}=\alpha\bm{r}$, $\bm{v'}=\bm{v}/\sqrt{\alpha}$, $t'=\alpha^{3/2}t$,  energy $E'=E/\alpha$ and angular momentum $L'=\sqrt{\alpha}L$. 
The known periodic orbits of the BHH family and their satellites with equal mass \cite{Broucke1975a, Henon1976, Jankovic2016} have zero total momentum (i.e.,  $m_1\dot{\bm{r_1}}+m_2\dot{\bm{r_2}}+m_3\dot{\bm{r_3}}=0$). 
Using the scaling law, we can transform the initial conditions of known periodic orbits of the BHH family and their satellites to the initial positions
\begin{equation}
\bm{r}_1(0)=(x_1,0), \;\; \bm{r}_2(0)=(1,0), \;\;  \bm{r}_3(0)=(0,0), 
\end{equation}
and the initial velocities
\begin{equation}   
 \dot{\bm{r}}_1(0)=(0,v_1), \;\; \dot{\bm{r}}_2(0)=(0,v_2), \;\; \dot{\bm{r}}_3(0)=\left(0, -\frac{m_1v_1+m_2v_2}{m_3}\right).
\end{equation} 

We use the numerical continuation method \cite{Allgower2003}  to gain the BHH orbits ($k=1$) and their satellites ($k>1$) with {\em unequal} mass.  Briefly speaking,  the  numerical continuation method can be used  to  gain periodic solutions of the nonlinear differential system
\begin{equation}
\dot{\bm{u}} = F(\bm{u},\lambda),
\end{equation} 
where $\lambda$  a  physical parameter, called ``natural parameter''.  Assume that $\bm{u}_0$ is a solution at a natural parameter  $\lambda = \lambda_0$.  Using the solution $\bm{u}_0$ at  $\lambda = \lambda_0$ as an initial guess, a new periodic orbit $\bm{u}'$  can be obtained at a new natural parameter $\lambda = \lambda_0 + \Delta \lambda$ through the Newton-Raphson method \cite{Farantos1995, Lara2002} and the clean numerical simulation (CNS) \cite{Liao2009,  Liao2014-SciChina,  Hu2020} if the increment $\Delta \lambda$ is small enough to make sure iterations convergence. 
The CNS is a numerical strategy to obtain reliable  numerical simulation of chaotic systems in a given time of interval.  The CNS is based on an arbitrary high order Taylor series method \cite{Chang1994, Barrio2005} and the multiple precision arithmetic \cite{Fousse2007}, plus a convergence check using an additional computation with even smaller numerical error.

Note that all of the known  BHH  orbits ($k=1$) and satellites ($k>1$) are ``relative periodic orbits'': after a period, these relative periodic orbits will return to initial conditions in a rotating frame of reference.  So,  there is an individual rotation angle
$\theta$ for each relative periodic orbit.  

First of all, using the known  BHH  orbits ($k=1$) and  satellites ($k>1$) with {\em equal} mass ($m_1=m_2=m_3=1$)  as initial guesses and $m_1$ as a natural parameter of the continuation method,
we obtain new periodic orbits with various  $m_1$ by continually correcting the initial conditions $x_1$, $v_1$, $v_2$, $T$ and the rotation angle $\theta$.   
Then, using these periodic solutions with $m_1\neq 1$,  $m_2=m_3=1$ as initial guesses and $m_2$ as a natural parameter of the continuation method,  we similarly gain periodic orbits for different values of $m_2$.   In this way,  we can obtain the corresponding  BHH (relative periodic) orbits ($k=1$) and satellites ($k>1$) with {\em unequal} mass $m_1\neq m_2\neq m_3 = 1$.

\begin{figure*}[]
\includegraphics[scale=0.25]{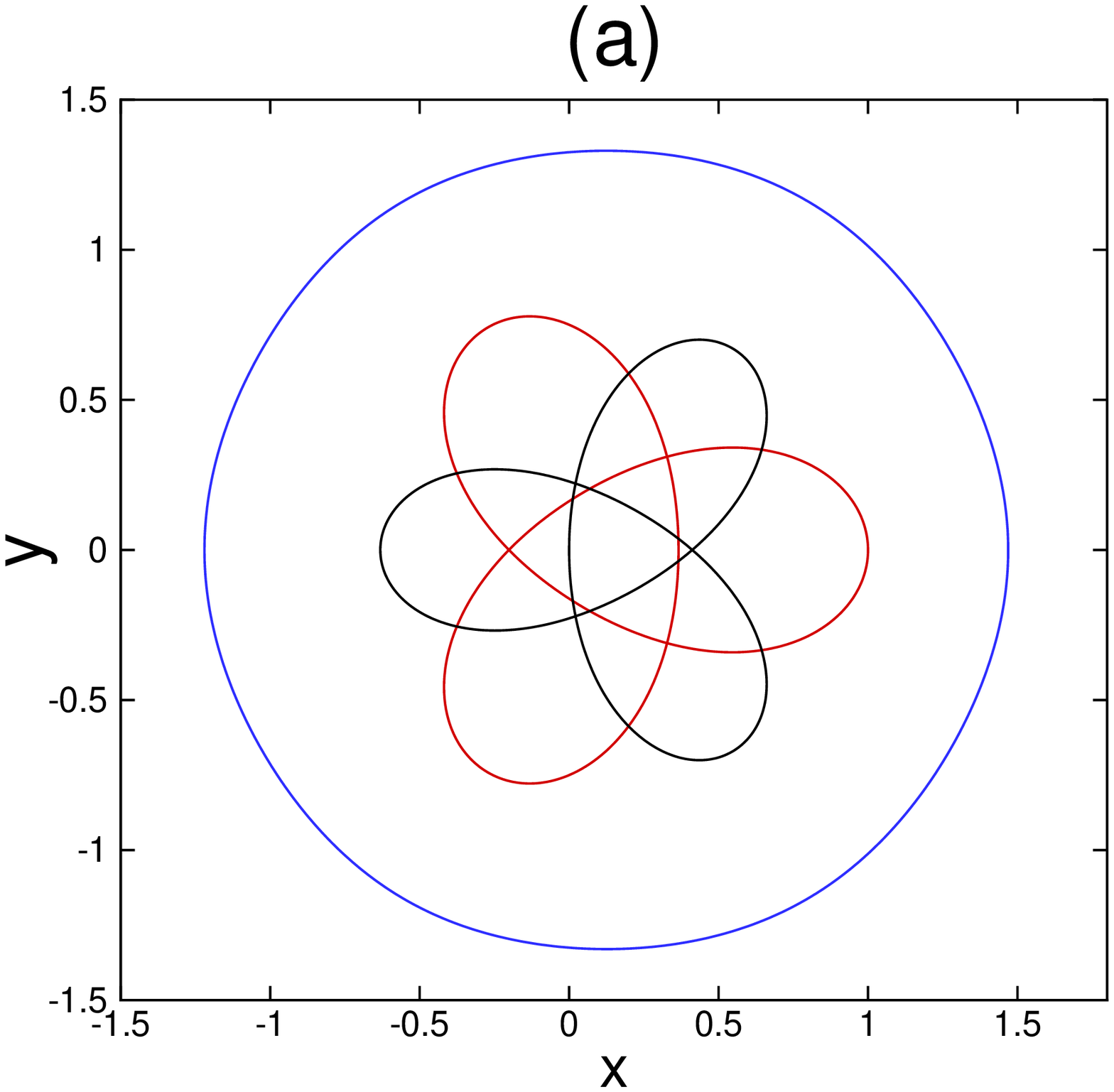}
\includegraphics[scale=0.25]{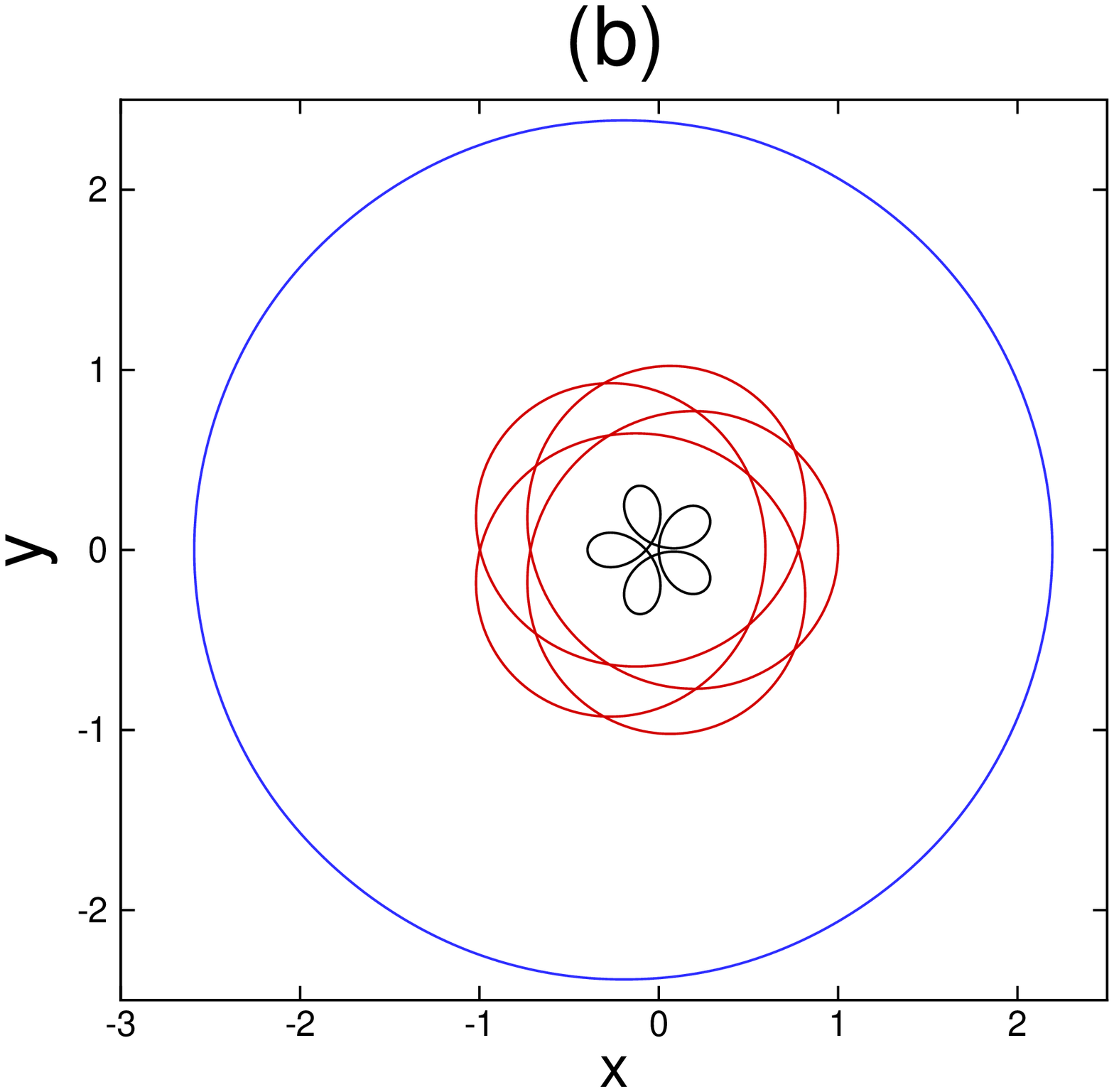}
\includegraphics[scale=0.25]{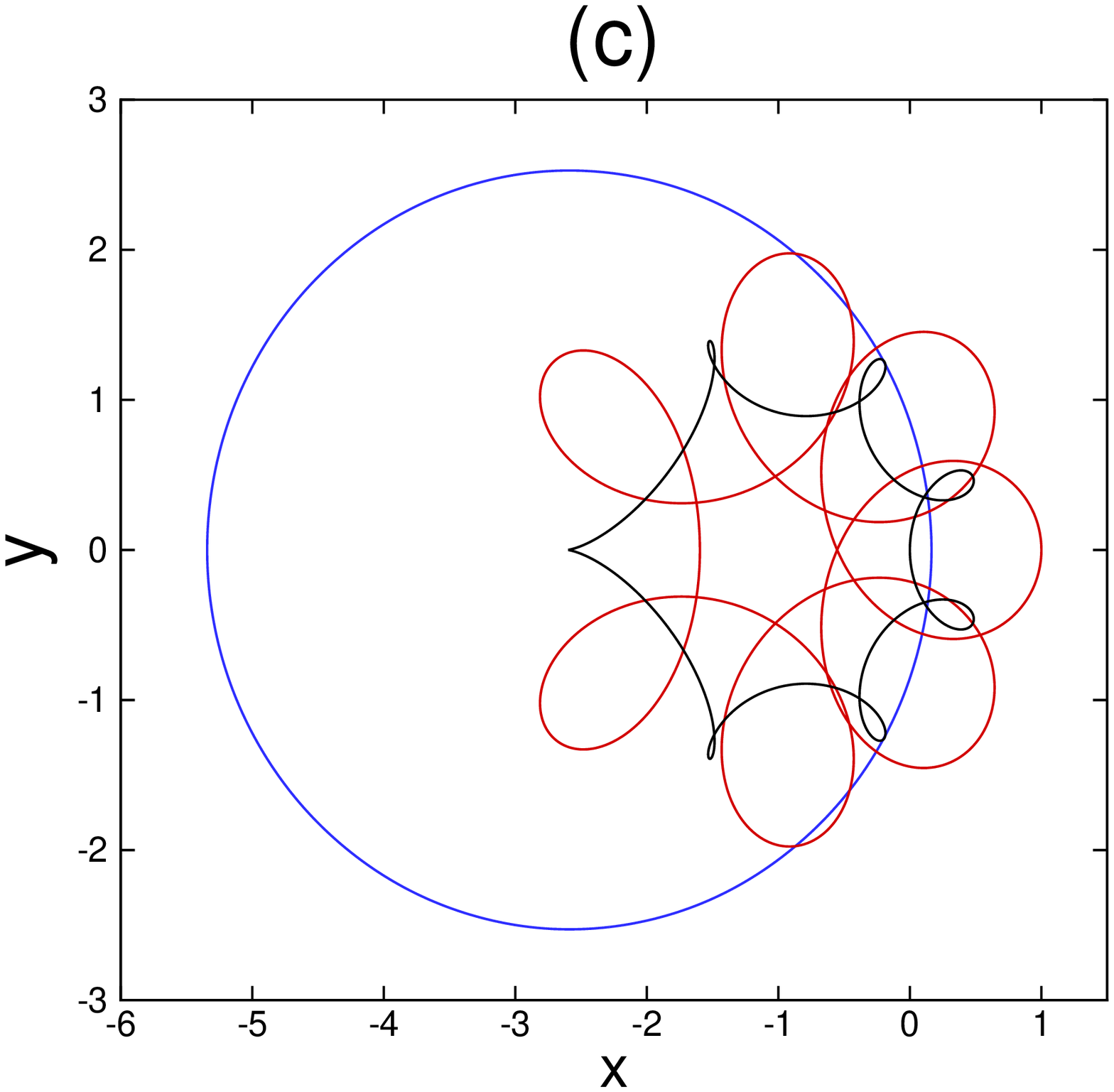}
\includegraphics[scale=0.25]{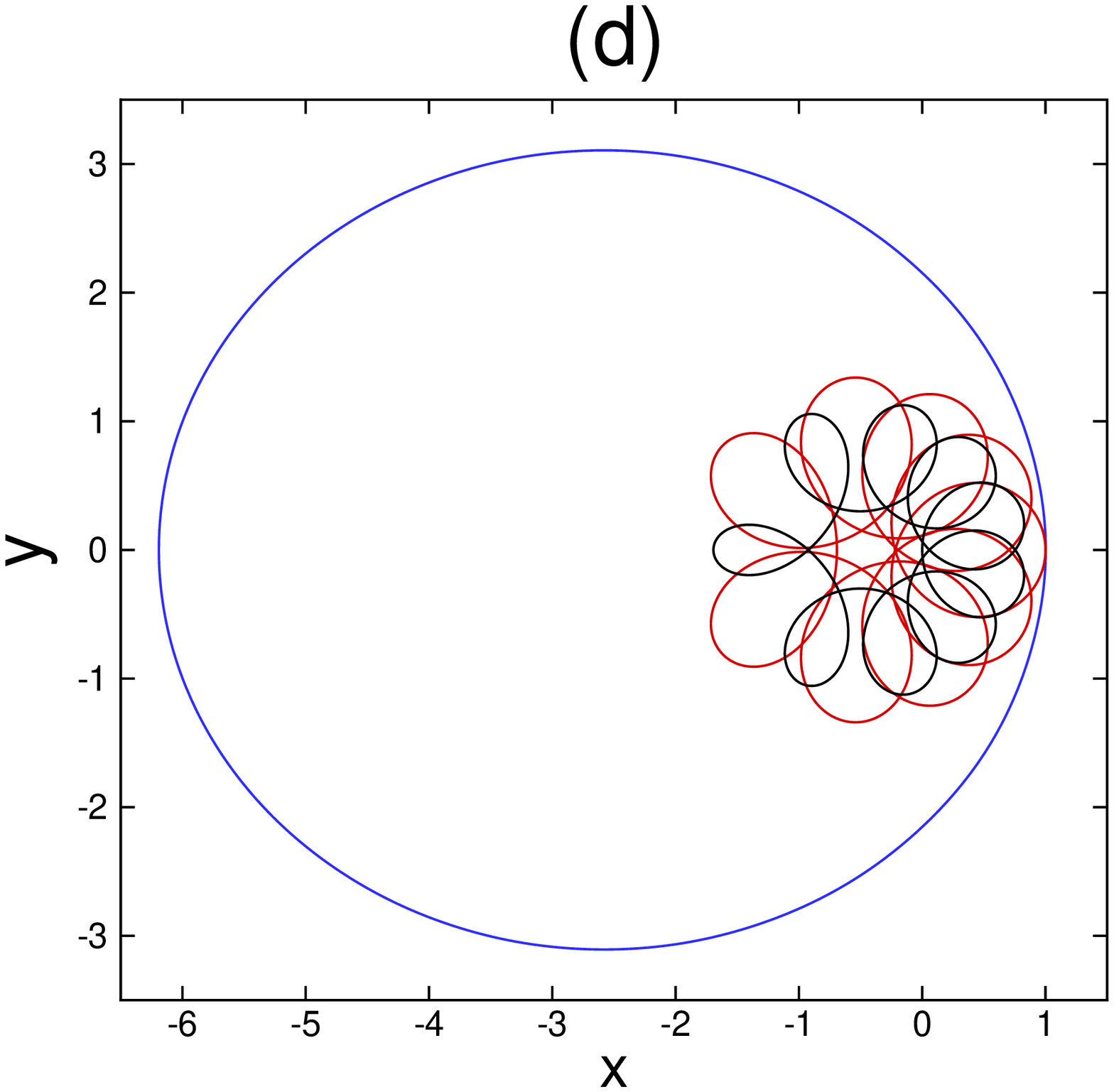}
\includegraphics[scale=0.25]{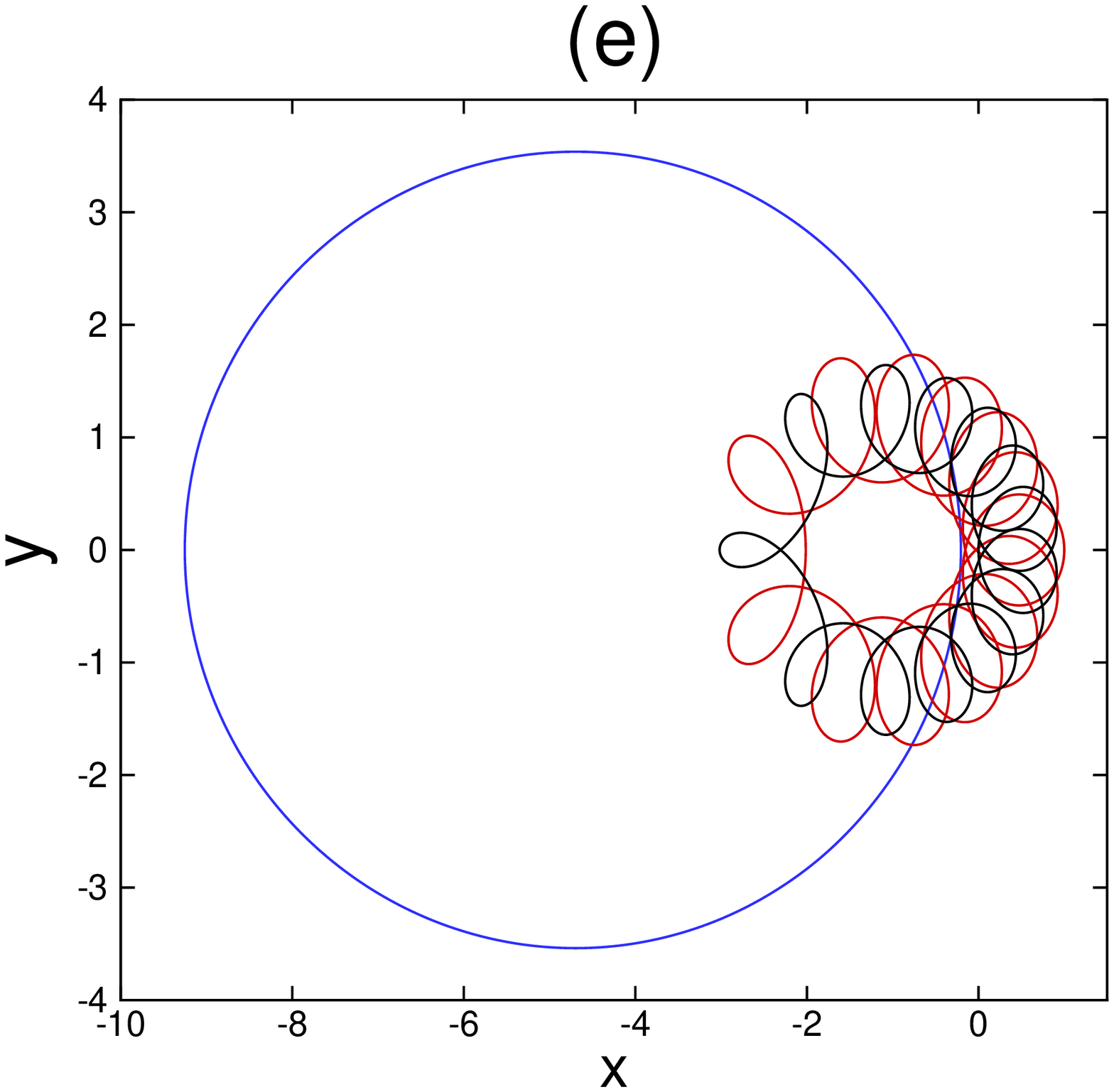}
\includegraphics[scale=0.25]{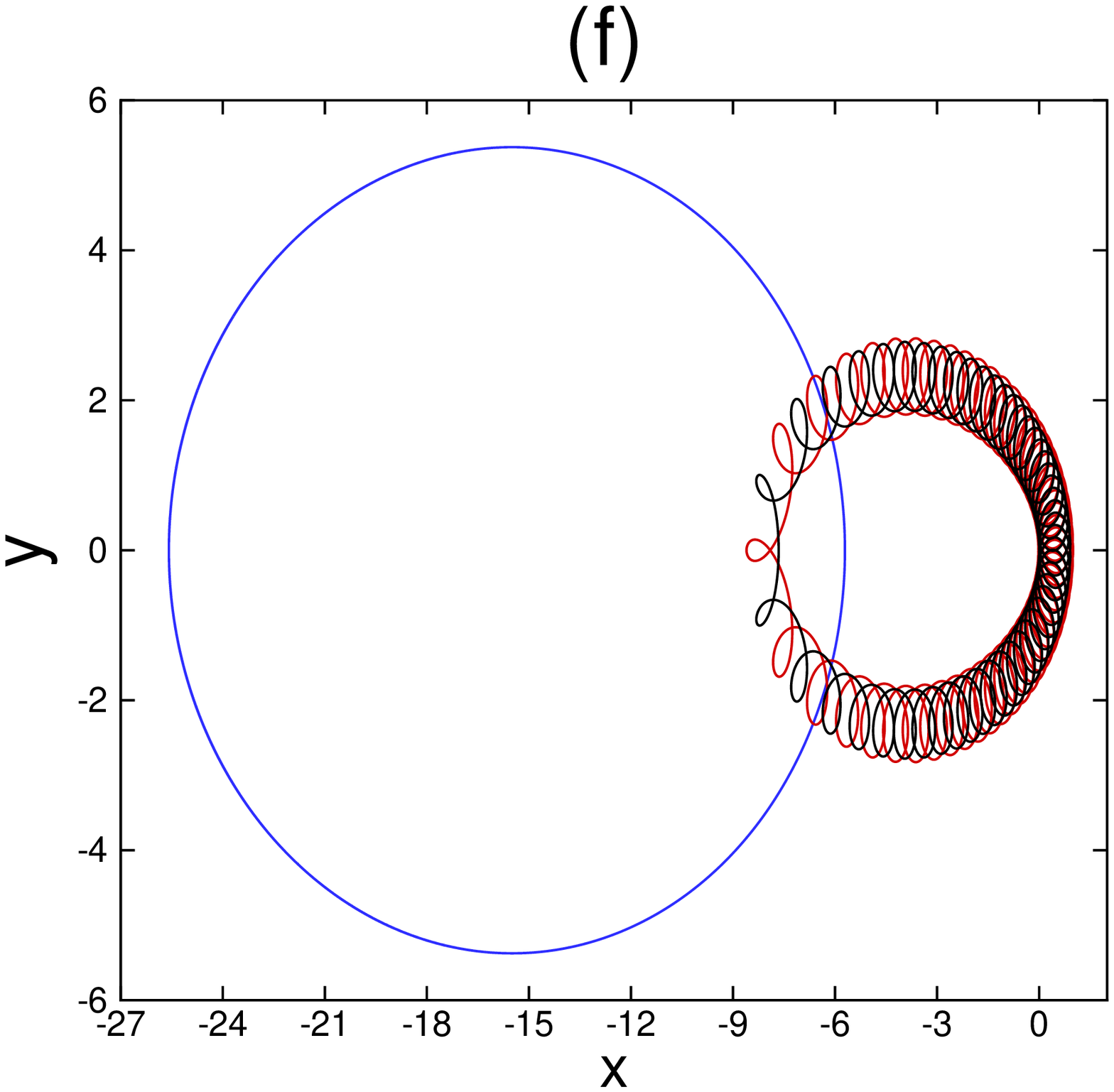}
  \caption{(color online.) The stable BHH satellites ($k>1$) of the three-body system with unequal mass in a rotating system.  Blue line: body-1; red line: body-2; black line: body-3.  The corresponding physical parameters are given in Table~\ref{ini}.}
  \label{orbit}
\end{figure*}

\begin{table}[h]
\tabcolsep 0pt \caption{Initial conditions and periods $T$ of some BHH satellites of three-body system  with unequal mass in case of  $\bm{r}_1(0)=(x_1,0)$, $\bm{r}_2(0)=(1,0)$, $\bm{r}_3(0)=(0,0)$, $\dot{\bm{r}}_1(0)=(0,v_1)$, $\dot{\bm{r}}_2(0)=(0,v_2)$,  $\dot{\bm{r}}_3(0)=(0,-(m_1v_1+m_2v_2)/m_3)$. Here $m_i$, $x_i$ and $v_i$ are the mass, initial position and velocity of the $i$th body,  $\theta$ is the rotation angle of relative periodic orbits, and $k$ is the topological power of periodic orbits, respectively.} \label{ini} \vspace*{-12pt}
\begin{center}
\def\temptablewidth{1\textwidth}
{\rule{\temptablewidth}{1pt}}
\begin{tabular*}{\temptablewidth}{@{\extracolsep{\fill}}cccccccccc}
No. & $m_1$ & $m_2$ & $m_3$ & $x_1$ & $v_1$  & $v_2$  & $T$ & $\theta$ & $k$\\
\hline
(a)	&	0.44	&	0.87	&	1	&	-1.21992948117021	&	-0.992252134619392	&	-0.513024298255905	&	9.1758282973000	&	0.500325594634030	&	3	\\
(b)	&	0.1	&	0.2	&	1	&	-2.59038883768724	&	-0.619538016547757	&	-0.865730420457027	&	23.1822105206534	&	0.110299604356735	&	5	\\
(c)	&	0.64	&	0.36	&	1	&	-5.34303779563735	&	-0.320961649402539	&	-0.737471717803478	&	36.3965789125352	&	0.085296971469750	&	7	\\
(d)	&	0.4	&	0.7	&	1	&	-6.19095126372906	&	-0.330208056009860	&	-0.703472163111217	&	40.8464178849168	&	0.107927979415387	&	9	\\
(e)	&	0.6	&	0.8	&	1	&	-9.25316717310693	&	-0.235487010686519	&	-0.683551074764106	&	60.3350581589708	&	0.086983638647642	&	13	\\
(f)	&	0.82	&	0.9	&	1	&	-25.5854144048497	&	-0.093120912929298	&	-0.673916187504190	&	204.731304275836	&	0.056232585939249	&	48	\\
\end{tabular*}
{\rule{\temptablewidth}{1pt}}
\end{center}
\end{table}

\begin{figure}[]
  \includegraphics[scale=0.4]{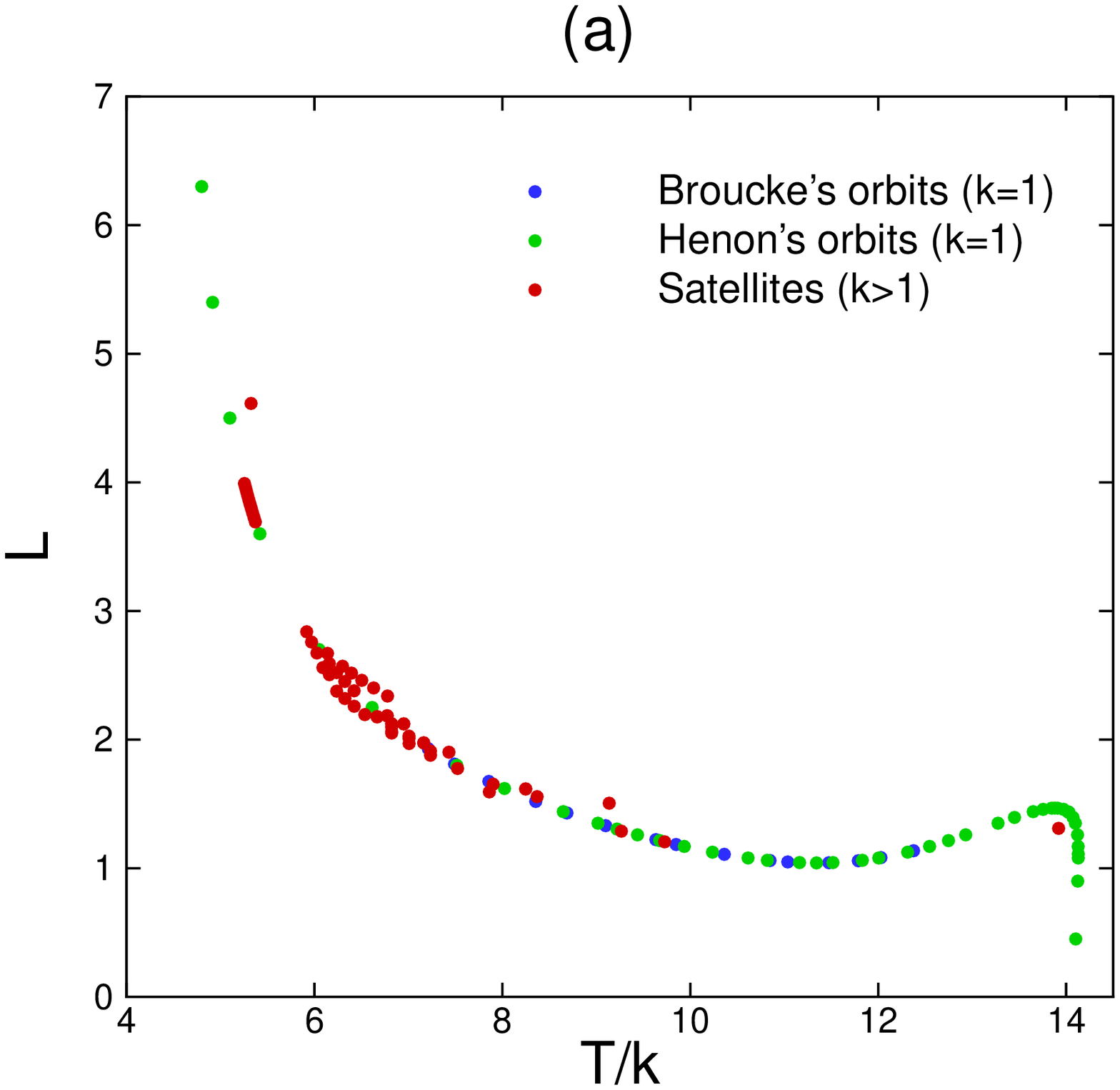}
  \includegraphics[scale=0.4]{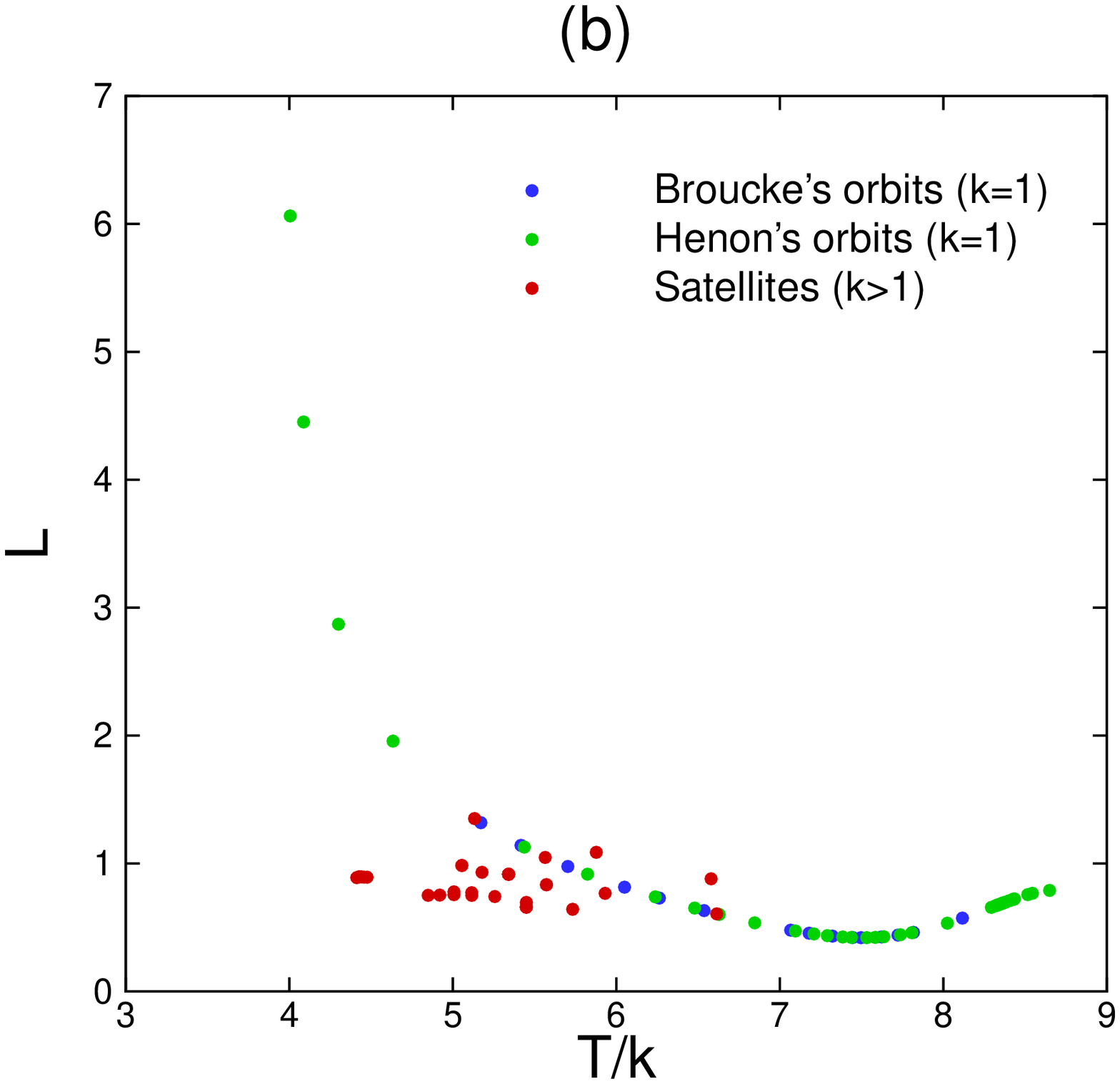}
  \includegraphics[scale=0.4]{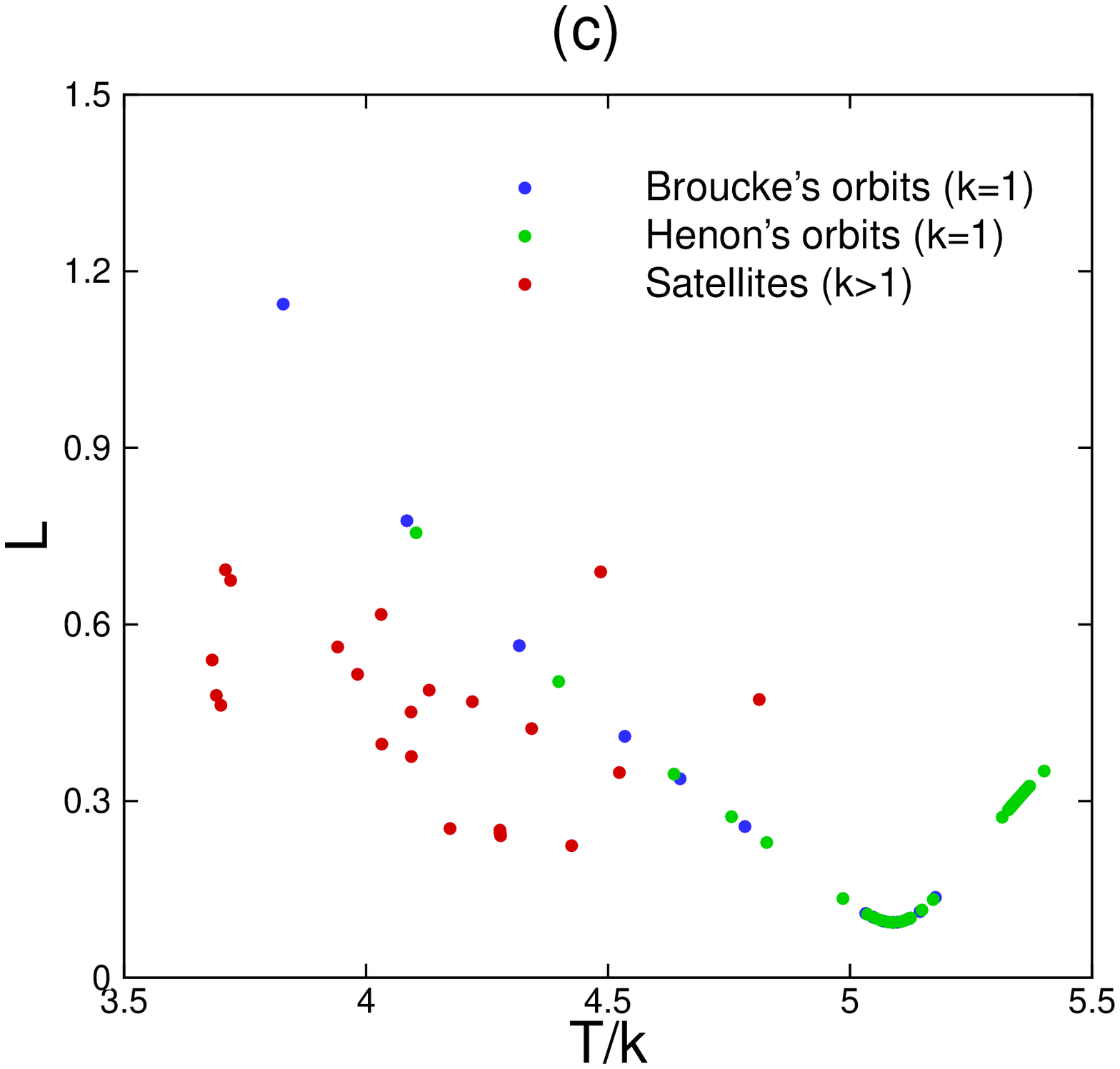}
  \includegraphics[scale=0.4]{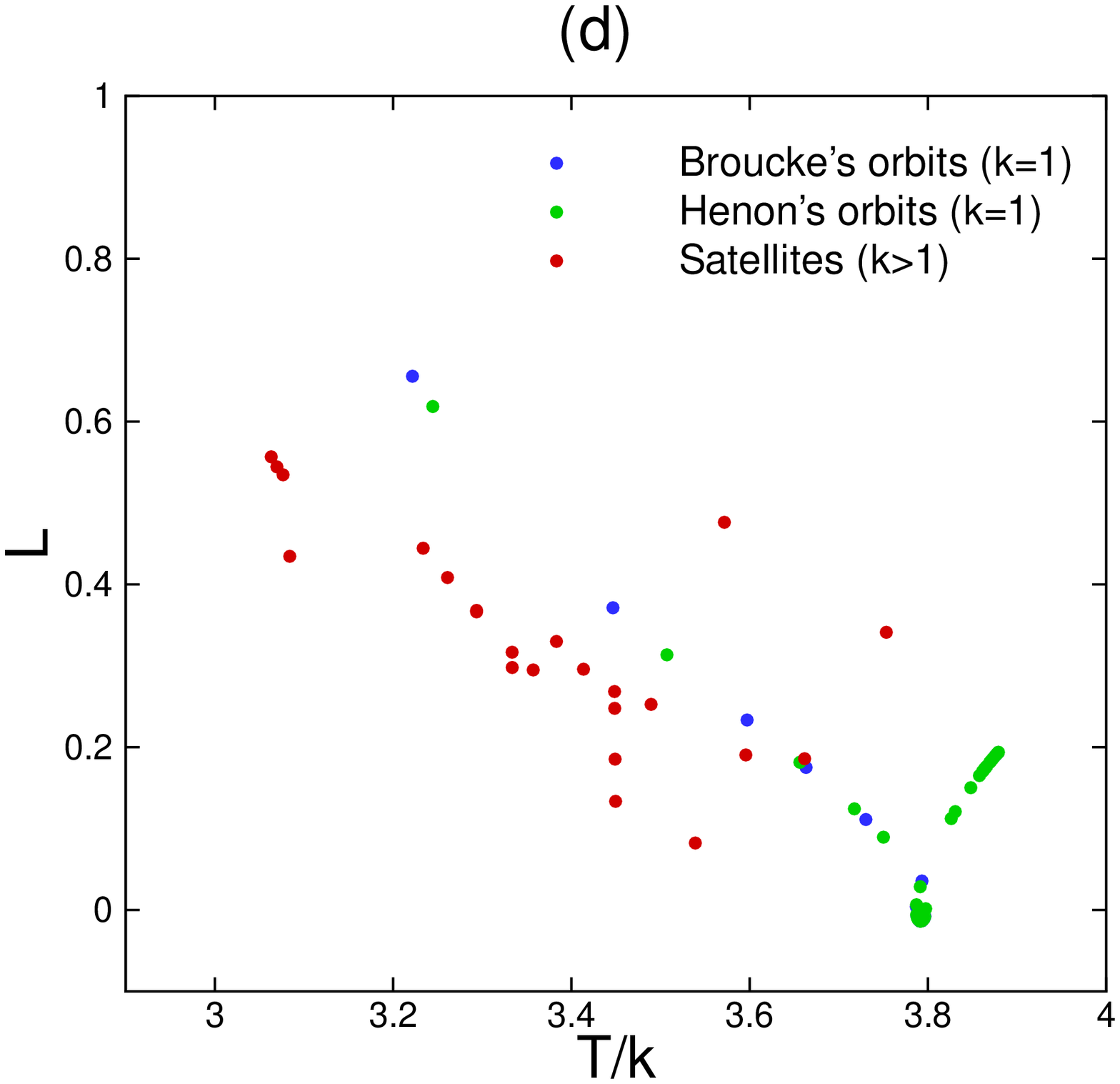}
  \caption{(color online.) The angular momentum ($L$) versus the topological rescaled period ($T/k$) for  BHH periodic orbits and their satellites at fixed energy $E=-1/2$ with different mass: (a) $m_1=m_2=m_3=1$; (b) $m_1=0.7$, $m_2=0.9$, $m_3=1$; (c) $m_1=0.5$, $m_2=0.8$, $m_3=1$; (d) $m_1=0.4$, $m_2=0.7$, $m_3=1$.}
  \label{L-T}
\end{figure}

\section{Results}

Starting from the 16 known Broucke's periodic orbits ($k=1$ with equal mass) \cite{Broucke1975a}, the 45 known  H\'{e}non's periodic orbits ($k=1$ with equal mass) \cite{Henon1976} and the 58 known BHH satellites ($k>1$ with equal mass) \cite{Jankovic2016},
we obtain  124780,  294963 and  179253 {\em new} periodic orbits of the three-body system with {\em unequal mass} for  $m_1 \in [0.1,1)$, $m_2 \in [0.1, 1)$ and $m_3=1$,  respectively.
Totally, we gain 419,743 new BHH  orbits ($k=1$ with {\em unequal} mass)  and 179,253 new BHH satellites ($k>1$ with {\em unequal} mass).   
Note that all of them are retrograde, say, the binary system and the third body move in opposite direction.     
The corresponding initial conditions, the periods and the rotation angles  of these new BBH orbits and satellites with unequal mass  are given in the supplementary data.    
The return distance of these periodic orbits and satellites satisfies 
\[  d=\sqrt{\sum_{i=1}^3((\bm{r}_i(T)-\bm{r}_i(0))^2+(\dot{\bm{r}}_i(T)-\dot{\bm{r}}_i(0))^2)} < 10^{-10}\]
 in a rotating frame of reference, where $T$ is the period.
Note that Broucke and Boggs \cite{Broucke1975b}  gave dozens of the BHH orbits with unequal mass (their ratios of mass are different from ours),  but {\em neither} have any BHH satellites with {\em unequal} mass been reported, to the best of our knowledge.   It should be emphasized that,  among our newly-found 598,996 BHH orbits and satellites with {\em unequal} mass, there are 151,925 stable BHH orbits ($k=1$)  and 48,761 stable BHH satellites ($k>1$).   In other words,  about 33.5\% of the new BBH orbits and satellites with {\em unequal} mass (i.e., 200,686) are stable,  and thus many among them might be observed in practice.   The stability of these periodic orbits and satellites is marked  by  ``S''  in the supplementary data.
Six new BHH satellites with {\em unequal} mass are shown in Figure~\ref{orbit}.   All of the six orbits are linearly stable.    Their initial conditions, periods and topological powers are listed in Table~\ref{ini}.  Note that these orbits are relatively periodic, say,  their orbits are closed curves in a rotating frame of reference.

With rescaling to the same energy $E=-1/2$, 
Jankovi\'{c} and Dmitra\v{s}inovi\'{c} \cite{Jankovic2016} found that, in case of {\em equal} mass,  the relationship between the scale-invariant angular momentum ($L$) sand  the topologically rescaled period ($T/k$) is the {\em same} for {\em both} of the BHH orbits ($k=1$) and satellites ($k>1$),  as shown in Figure~\ref{L-T}(a), where $k$ is the topological exponent of periodic orbits.    
However,  for our newly-found periodic orbits with {\em unequal} masses (at the same energy $E=-1/2$),  it is found that the relationship between the scale-invariant angular momentum ($L$) and period ($T/k$) of the BHH satellites ($k>1$) is different from that of the BHH orbits ($k=1$),  as illustrated in Figure~\ref{L-T}~(b)-(d).
It suggests that the relationship between the scale-invariant angular momentum ($L$) and topologically rescaled period ($T/k$) of the BHH orbits ($k=1$) and satellites ($k>1$) in general cases of {\em unequal} masses $m_1\neq m_2\neq m_3$ should be more complicated than that in the case of the equal mass $m_1=m_2=m_3$.    

\section{conclusion}

The BHH orbits are a family of periodic orbits of the three-body system with the simplest topological free group word $a$, while the BHH  satellites have free group words $a^k$ ($k>1$), where $k$ is the topological exponent.   
In this paper,  starting from the 16 known  Broucke's periodic orbits ($k=1$ with {\em equal} mass) \cite{Broucke1975a}, the 45 known  H\'{e}non's periodic orbits ($k=1$ with {\em equal} mass) \cite{Henon1976} and the 58 known BHH satellites ($k>1$ with {\em equal} mass) \cite{Jankovic2016}, we found 419,743 new BHH orbits ($k=1$) and 179,253 new BHH satellites ($k>1$) for three-body system with {\em unequal} mass,  which have never been reported,  to the best of our knowledge.    Among these newly-found 598,996 BBH  orbits and satellites of three-body system with unequal mass,  about 33.5\% (i.e.,  200,686) are linearly stable and thus many among them might be observed in practice. 

For the three-body system with {\em equal} mass at a fixed energy $E=-1/2$,   it was reported \cite{Jankovic2016}  that the relationship between the angular momentum ($L$) and topological period ($T/k$) of the BHH satellites ($k>1$) is the same as that of the BHH orbits ($k=1$).   However, this does not hold for the three-body system with {\em unequal} mass, as reported in this letter.  

This work was carried out on TH-1A at National Supercomputer Center in Tianjin and TH-2 at National Supercomputer Center in Guangzhou, China.  It is partly supported by National Natural Science Foundation of China (Approval No. 91752104) and the International Program of Guangdong Provincial Outstanding Young Researcher.


\bibliography{ref}

\end{document}